# Superconductivity above 180 K in Ca-Mg Ternary Superhydrides at Megabar Pressures

Weizhao Cai[1,2], Vasily S. Minkov[2], Ying Sun[3], Panpan Kong[2], Krista Sawchuk[4], Boris Maiorov[4], Fedor F. Balakirev[4], Stella Chariton[5], Vitali B. Prakapenka[5], Yanming Ma[3], Mikhail I. Eremets[2,*]

[1]School of Materials and Energy, University of Electronic Science and Technology of China, Chengdu 611731, Sichuan, China

[2]Max Planck Institute for Chemistry, Hahn Meitner Weg 1, Mainz 55128, Germany

[3]State Key Laboratory of Superhard Materials and International Center for Computational Method & Software, College of Physics, Jilin University, Changchun 130012, China

[4]MagLab, Pulsed Field Facility, Los Alamos National Laboratory, Los Alamos, NM 87545, USA

[5]Center for Advanced Radiation Sources, University of Chicago, Chicago, IL 60637, USA

*E-mail: m.eremets@mpic.de

The discovery of high-temperature superconductivity above 240 K in binary La-H and Y-H systems inspired further predictions of even higher transition temperatures in compounds such as $YH_{10}$ and $MgH_6$, which are likely to be dynamically unstable. Ternary superhydrides provide alternative pathways to stabilize desired near-room temperature superconducting phases. However, the synthesis of new ternary hydrides remains challenging because most of the precursor reactants do not exist in desired stoichiometry at ambient conditions. Here we report that using the existing binary intermetallic $CaMg_2$ and 1:1 Ca-Mg mixture as starting reactants, we have successfully synthesized novel Ca-Mg-based ternary superhydrides at megabar pressures. Electrical resistivity measurements show $T_c$ approaching 168 K at 310 GPa in the $CaMg_2$-based superhydride and 182 K in 1:1 the Ca-Mg superhydride at 324 GPa.

The observation of a high superconducting critical temperature ($T_c$) of ~203 K in $H_3S$ has prompted further exploration of other high $T_c$ hydrogen-rich compounds based on Ashcroft's hydrogen "chemical precompression" concept.[1] Theoretical simulations of phase diagrams of binary metal superhydrides have led to the discovery of new superconducting materials at megabar pressures. Unlike the H atoms in $H_3S$, which are covalently bonded in a three-dimensional framework, the structures of the newly discovered high-$T_c$ compounds generally exhibit a clathrate-cage-like hydrogen network.[2, 3] The rare earth and alkaline earth superhydrides are prominent examples of these systems. The sodalite-like clathrate $Fm\bar{3}m$-$LaH_{10}$ was predicted to have $T_c$ of ~280 K at 210 GPa,[4] and this was confirmed by subsequent experiments with $T_c$ = 250-260 K at ~180 GPa.[5-7] The finding of record $T_c$ in $LaH_{10}$ has stimulated more experimental explorations of Y-H system, since the $Fm\bar{3}m$-$YH_{10}$ stoichiometry is predicted to be a room temperature superconductor ($T_c$ of ~303 K at 400 GPa).[8-10] Intensive efforts have been spent to synthesize this compound, but no appearance of $YH_{10}$ up to 410 GPa and 2250 K has been observed. Only the low stoichiometries $Im\bar{3}m$-$YH_6$ and hexagonal close packed (*hcp*) $P6_3/mmc$-$YH_9$ have been synthesized and they show maximum $T_c$ of ~220 K (at 180 GPa) and 243 K (at 201 GPa), respectively, which are ~30 K lower than the calculated $T_c$ values.[11, 12]

The phase diagrams of most binary hydrides have been studied by theoretical simulations,[3, 13] and recent investigations have been extended to ternary systems, such as fluorite-type $AXH_8$ family,[14-16] $CaRH_{12}$ ($R$ = Y, Zr and Hf),[17, 18] and Mg-Sc/Y-H system[19, 20]. One prominent example is the clathrate structure $Li_2MgH_{16}$. The doped lithium atoms introduced extra electrons into the molecular-like hydrogen in the parent compound $MgH_{16}$, leading to a strong electron-phonon coupling and resulting in a $T_c$ of ~473 K at 250 GPa.[21] Hydrogen-rich ternary hydrides $Li_5MoH_{11}$[22] and $BeReH_9$[23] have been experimentally synthesized, but with low $T_c$ values (< 10 K), which is linked to the formation of the $H_2$ or $H_3$ molecular units within their structures that negatively affects the electronic density of states and the electron-phonon coupling. Recently, a ternary nonstoichiometric alloy (La,Y)$H_{10}$ superhydride with ~25% of the La atoms replaced by Y was synthesized with a $T_c$ of ~253 K at 183 GPa.[24] The $T_c$ value is very close to that of pure $LaH_{10}$ at comparable pressures.[5] The cage-like $P6_3/mmc$-$CeH_9$ exhibits low $T_c$ of 95–115 K around the one megabar pressure.[25] If half of the Ce atoms are randomly substituted by La atoms, the alloy (La,Ce)$H_9$ demonstrates 80% increase in $T_c$ compared to pure $CeH_9$.[26] Therefore, chemical substitution is an efficient route to tune the $T_c$ at high pressures.[27] Since there are no binary La-Y and La-Ce intermetallics exist in the binary phase diagram at ambient conditions,[28] the specific ratios of mixed metals with good homogeneity should be carefully prepared and checked before loading the diamond anvil cell when the micron-sized sample is employed as a reactant in the cell chamber.

The sodalite-like clathrate $Im\bar{3}m$-$CaH_6$ was first theoretically predicted to be a high-temperature superconductor with $T_c$ of 220-235 K at 150 GPa in 2012,[29] and recently confirmed by two independent groups.[30, 31] The prediction of high $T_c$ in $CaH_6$ inspired further studies on the isotypic hexahydride $Im\bar{3}m$-$MgH_6$, which was predicted to be a conventional Bardeen-Cooper-Schrieffer (BCS) superconductor with $T_c$ = 260 K at 300 GPa.[32] When Ca is substituted by 50% of Mg, the substitutional alloy $Im\bar{3}m$-$Ca_{0.5}Mg_{0.5}H_6$ is predicted to be a near-room-temperature superconductor with $T_c$ of 288 K at 200 GPa, which is much higher than that of its parent compounds $MgH_6$ or $CaH_6$.[33, 34] Other high stoichiometry calcium hydrides such as $C2/m$-CaH$_9$ and $R\bar{3}m$-CaH$_{10}$ are predicted to be high-$T_c$ superconductors with $T_c$ in the range of 240-266 K and 157-175 K at 300 and 400 GPa, respectively[35], while the high H-content magnetism-based hydrides $R3$-MgH$_{12}$ and $P\bar{1}$-MgH$_{16}$ exhibit low $T_c$ of 47-60 K at 140 GPa.[36] Experimentally, using a 1:1 ratio of Ca-Mg alloy as the reactant to synthesize this substituted material is challenging because the Ca$_1$Mg$_1$ intermetallic does not exist in the Ca-Mg phase diagram at ambient pressure, so the composition of the precursor should be carefully confirmed prior to sample loading. These experimental limitations emphasize that alternative starting reactants should be explored in the synthesis of novel high-$T_c$ superhydrides at high pressures. In the binary phase diagram of Ca-Mg, only the CaMg$_2$ intermetallic exists.[37] Previously reported synthesized Ca-Mg based ternary hydrides include $P\bar{6}2m$-Ca$_4$Mg$_3$H$_{14}$,[38] $Im\bar{3}m$-Ca$_{19}$Mg$_8$H$_{54}$[39] and CaMgH$_{3.72}$.[40] Earlier experiments reveal that CaMg$_2$ reacts with hydrogen (H$_2$) with reaction path: CaMg$_2$ + H$_2$ → CaH$_2$ + 2Mg at 450 °C.[41] If excess hydrogen present, CaMg$_2$ absorbs 4.48wt% of hydrogen at 573 K and transforms to Ca$_4$Mg$_3$H$_{14}$ and CaH$_2$ under 3.8 MPa of hydrogen.[42]

In this work, we present the synthesis of new ternary Ca-Mg-based superhydrides at megabar pressures. From high pressure synchrotron X-ray diffraction and electrical transport measurements, we find that synthesized CaMg$_2$-based superhydride adopts face-centered cubic (*fcc*) symmetry, which may have a symmetry similar to reported ternary superhydride $Fm\bar{3}m$-LaBeH$_8$. Such structure supports high temperature superconductivity above 100 K, with $T_c$ approaching ~168 K when pressure increases to 310 GPa. Interestingly, the superconducting $T_c$ increases with decreasing proportion of the Mg element in the synthesized ternary Ca-Mg-H superhydrides: using the 1:1 Ca-Mg alloy as the starting reactant, we find that the $T_c$ of the resulting superhydrides is higher than that of the CaMg$_2$-based superhydride at comparable pressures, and it reaches 182 K at 324 GPa.

**Results and Discussion**

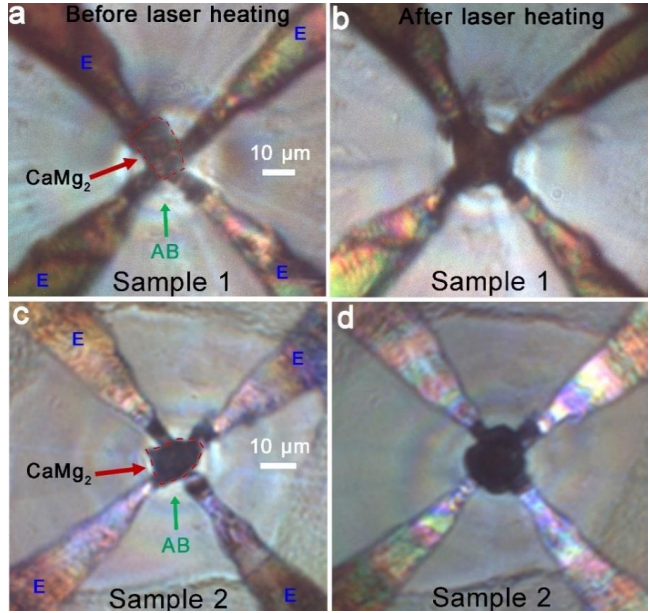

**Figure 1. Optical images of the CaMg$_2$ and ammonia borane before and after laser heating.** Panels (**a**) and (**b**) show the loaded sample before and after laser heating. The sample became semi-transparent after heating is performed. The pressure dropped from 192 to 185 GPa after laser heating is conducted. (**c**) and (**d**) are images of the Sample 2. The pressure remains the same ~215 GPa before and after laser heating. The CaMg$_2$ sample is marked by the dashed area and the letter "E" represents the electrode in all the images.

**Synthesis of the ternary Ca-Mg-based superhydrides**

To investigate the formation of ternary Ca-Mg hydrides and their superconducting properties, we loaded five diamond anvil cells (DACs) with sputtered electrodes using CaMg$_2$ alloy and ammonia borane (NH$_3$BH$_3$, AB) as the reactants. In addition, we have loaded an additional DAC with 1:1 Ca:Mg ratio material to study the effect of the Ca/Mg compositions on the superconducting properties of the ternary superhydride. The purity of the alloys was checked by the energy dispersive spectroscopy (EDS) prior to the sample loading (Figures S1 and S2). The mixtures of Ca-Mg alloy and AB were initially compressed to 180-260 GPa at room temperature and heated to 1200-2000 K via laser heating. We collected synchrotron X-ray diffraction data in all six DACs, and obtained X-ray data from three of them successfully, despite that elemental Ca and Mg are not good X-ray scatterers. Since the maximum $T_c$ ~210 K in the binary $CaH_6$ was observed at ~180 GPa, our first experiment, Sample 1, was compressed to ~185 GPa at room temperature.[31] Unexpectedly, the shiny $CaMg_2$ alloy became semi-transparent after laser heating to ~1200 K, and the pressure dropped slightly to 183 GPa (see Figure 1a). The collected X-ray diffraction (XRD) pattern of the annealed Sample 1 can be indexed as face-centered-cubic (*fcc*) lattice with lattice parameter $a = 4.6920(5)$ Å and $V = 103.29(1)$ Å$^3$ (see Figure 2a and Table S1). The cubic *fcc* symmetry structure of the synthesized sample is most likely similar to the reported binary H-rich superconductors $Fm\bar{3}m$-$LaH_{10}$,[5, 6] and $Fm\bar{3}m$-$LaBeH_8$[43]. The electrical resistance data revealed the synthesized sample shows semiconducting upturn with decreasing temperature even when high heating temperature of ~1600 K is used, and in addition we found the XRD patterns is very similar to that of ~1200 K heated sample, as we discuss below. Therefore, we suspect that the pressure used in synthesis of Sample 1 is not high enough to form the superconducting phase, so subsequent experiments with pressure exceeding two megabars were conducted.

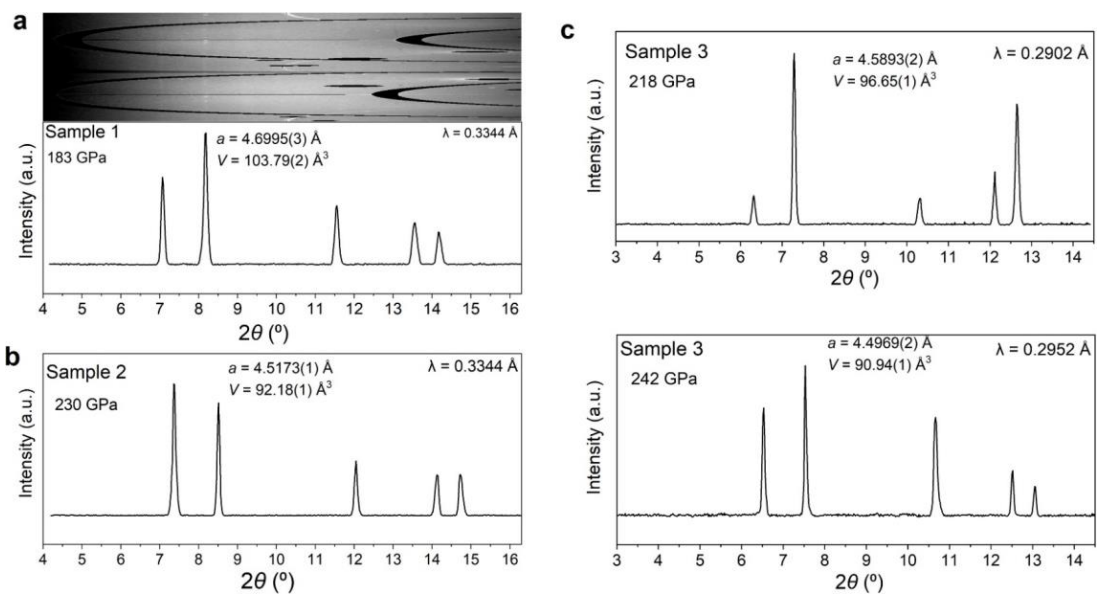

**Figure 2. Synchrotron X-ray data of synthesized samples from $CaMg_2$ alloy and ammonia borane measured at room temperature.** (**a**) The experimental XRD pattern for Sample 1 at 183 GPa (heated at 1200-1600 K) and (**b**) Sample 2 at 230 GPa (heated at ~1600 K) and (**c**) Sample 3 at 218 GPa and 242 GPa (heated at ~1700 K). Three heating cycles were performed under compression, which may cause the obvious preferred orientations of the XRD patterns between 218 and 242 GPa. Note that the XRD background signal has been subtracted in all panels.[44]

Two DACs, Samples 2 and 3, were prepared and the XRD data were collected at 230 and 242 GPa, respectively. After the mixture of $CaMg_2$ and AB laser heated at ~1500-1700 K in Sample 2 and 3, the apparent volume expansion is observed, and the color changes to black (see Sample 2 in Figure 1b and Table S1). The XRD data revealed that the synthesized materials in both DACs crystallizes in the *fcc* symmetry, with the lattice parameter *a* equal to 4.5103 (5) and 4.5892(1) Å for pressure at 230 and 218 GPa, respectively (Table S1). After the powder X-ray data for Sample 3 at 218 GPa were collected, we increased the pressure to 242 GPa and a few cycles of heating at ~1700 K were conducted. We observed the XRD patterns of sample shows preferred orientations of between 218 and 242 GPa (Figure 2c). The fitted XRD pattern yields $a$ = 4.4969(2) Å at 242 GPa. The hydrogen stoichiometry can be estimated from the molecular volume of $CaMg_2$ and the volume per one hydrogen atom. From the calculated equation of states of $CaMg_2$,[45] the corresponding formula-unit volume is approximately 30.6. The volume of one hydrogen atom estimated from the volume of the $CaH_6$ hydride is roughly 1.7 Å$^3$ at 242 GPa, so the hydrogen number $n$ can be calculated as $n = [V_{mol}(sample) − V_{mol}(CaMg_2)]/V_H = [(V_{cell}(sample)/2 − V_{cell}(CaMg_2))/4)]/V_H = (45.47 − 30.4)/1.7 = 8.86$, which is slightly larger than the ideal stoichiometry of 8. The obtained formula of the synthesized hydrides comparable with a recently predicted fluorite-type ternary structure $Fm\bar{3}m$-AXH$_8$ ($Z$ = 4).[16] Calculations of the stable $CaMg_2$-based superhydrides in the Ca-Mg-H ternary system are in progress.

**Superconductivity of synthesized superhydrides**

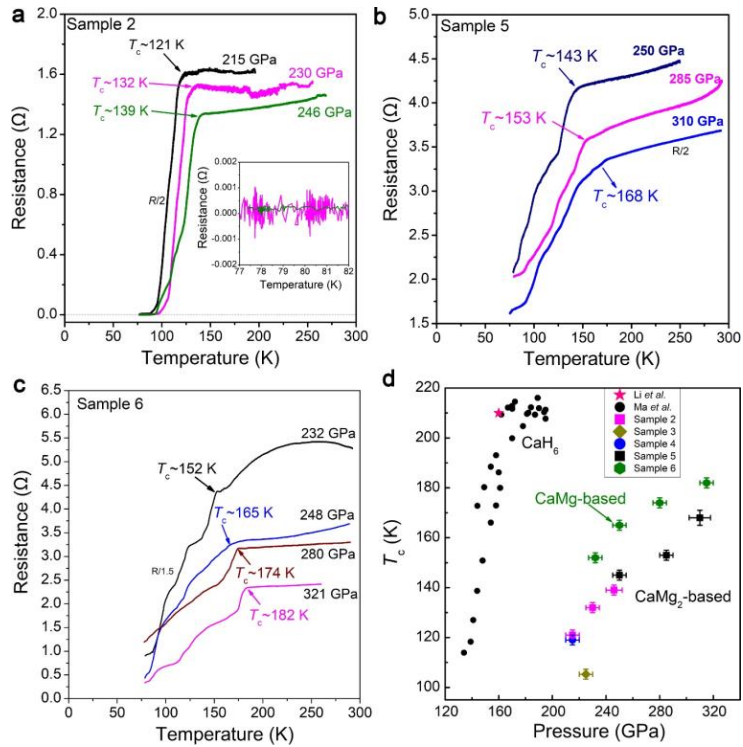

**Figure 3. Superconducting transitions of the synthesized Ca-Mg hydrides.** (**a**) The electrical resistance measurements of Sample 2 using the four-probe van del Pauw geometry. (**b**) The temperature dependence of electrical resistance was measured for Sample 5. (**c**) Superconducting transitions of synthesized Sample 6 prepared from the mixture of 1:1 Ca-Mg and ammonia borane. The mixture was laser heated at ~1600 K and compressed to 321 GPa isothermally (at room temperature). The critical temperature $T_c$ shifts to a higher temperature under compression. (**d**) The pressure dependence of $T_c$ from all experiments using CaMg$_2$ and CaMg as the reactants. The measured $T_c$s from five different experiments are shown in different symbols. Data from CaH$_6$ (refs. 25 and 26) are added for comparison.

We performed electrical resistance measurements using van der Pauw 4-wire topology to detect the superconductivity in the synthesized samples. As discussed above, the synthesized Sample 1 at 183 GPa which was heated to 1200-1600 K shows semi-transparent color and the $R(T)$ curve shows negative $dR/dT$ in the range of 80-300 K, indicating semiconducting character (Figure S3). Subsequent DACs were prepared above two megabars to explore the emergence of superconductivity. As shown in Figure 3a, the $R(T)$ curves from Sample 2 show metallic character prior to the appearance of the sharp drops, with clear superconducting transitions observed at ~121, 132 and 139 K for pressures $p$ = 215, 230 and 246 GPa, respectively. Zero-resistance states were also observed at 230 and 246 GPa in Sample 2 (see inset in Figure 3a). The critical superconducting transition temperature $T_c$ in the CaMg$_2$-based ternary superhydride is lower than the peak $T_c$ reported for the binary cubic $Im\bar{3}m$-CaH$_6$ alloy,[30, 31] and the $T_c$ predicted for Ca$_{0.5}$Mg$_{0.5}$H$_6$.[33] To investigate the pressure dependence of $T_c$, we have loaded Sample 4 and 5 into two DACs and compressed them to the maximum pressure of 310 GPa. The $T_c$ values increase monotonically, with a highest observed $T_c$ of ~168 K at 310 GPa (Figure 3d). We assume that if further compression were conducted, a higher $T_c$ would be expected. Moreover, we also prepared Sample 6 from the 1:1 mixture of Ca-Mg (CaMg) and AB. Interestingly, the measured $T_c$ at 232 GPa is ~152 K, which is much higher than that of the sample prepared from the $CaMg_2$ at comparable pressure. The critical temperature $T_c$ increased to ~182 K when the pressure reached 321 GPa (Figure 3c). The pressure required to synthesize the superconducting phase increases dramatically with increasing Mg content, starting with 130 GPa for pure $CaH_6$ and increasing to 215 GPa for $CaMg_2H_x$, while the superconductivity in 1:1 Ca:Mg hydride require about 30 GPa less pressure than $CaMg_2H_x$ to produce similar $T_c$ values.

**Electrical resistance measurements under various magnetic field**

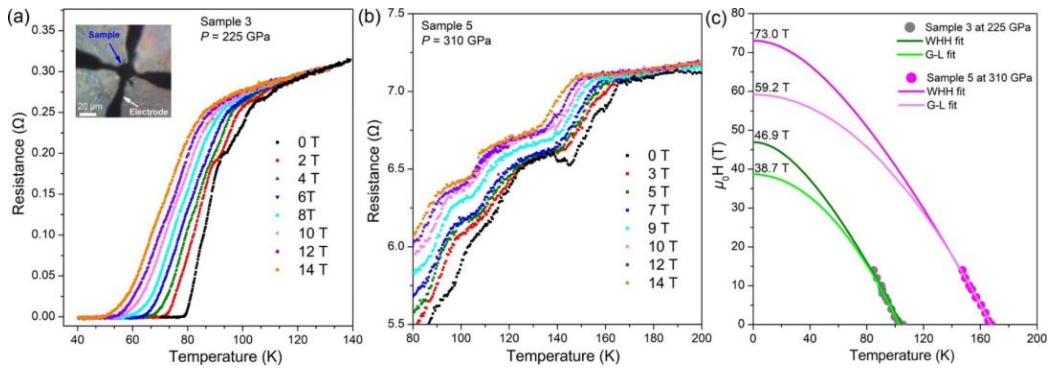

**Figure 4. Temperature dependence of electrical resistance of *fcc* $CaMg_2$-based superhydride under external magnetic field.** (a) Temperature dependence of electrical resistance of Sample 3 under external magnetic field of 0-14 T at 225 GPa. The inset indicates the sample image under transmission light. (b) *R-T* dependence of Sample 5 under external magnetic field at 310 GPa. (c) Fittings of the upper critical field $H_{c2}$ of sample 3 and 5 using Werthamer-Helfand-Hohenberg (WHH) and Ginzburg-Landau (GL) models.

We have performed magneto-transport measurements under the external magnetic field up to 14 T to verify the presence of the superconductivity in the synthesized $CaMg_2$-based superhydrides. Magnetic field suppresses the superconducting phase, and the superconducting transition temperature decreases with increased the magnetic field. We find that the suppression of superconducting magnetic field is consistent with previously reported superhydrides.[5, 11] Figures 4 a and 4b show the superconducting transitions in the Samples 3 and 5 at different magnetic field values, respectively. We determine the key superconducting properties, including upper critical field $H_{c2}$ and superconducting coherence length $\xi = \sqrt{\frac{\phi_0}{2\pi H_{c2}}}$, where $\Phi_0$ is the magnetic flux quantum (Table 1). For sample 3 at 225 GPa, the upper critical field $H_{c2}(0)$ estimates using the Werthamer-Helfand-Hohenberg (WHH) and Ginzburg-Landau (GL) models reaches 46.7 T and 38.7 T, respectively. For Sample 5 at 310 GPa, the estimated $H_{c2}$ for WHH and GL models equals 73.0 T and 59.2 T, respectively (Figure 4c). The extrapolated $T \to 0$ upper critical field is much lower than the one reported for the binary high temperature superconductors above 200 K, e.g. $T_c$ = 215 K and $H_{c2}^{WHH}$ = 203 T for $Im\bar{3}m$-$CaH_6$ at 172 GPa.[31] The coherence length and the $T_c$ in a superconductor are linked to the carrier Fermi velocity: $v_F = \xi k_B T_c / 0.18\,\hbar$,[46] which is about $2.5\times10^5$ m/s in $CaMg_2H_x$, indicative of the Fermi velocity renormalization due to strong electron-phonon pairing, as it was suggested by very similar Fermi velocities observed in most of the high-$T_c$ hydrides (Table 1 and Figure S5).[47]

Table 1. Summary of the synthesized sample 3 and 5 and the associated WHH fit parameters.

| Sample | Pressure (GPa) | $T_c$ (K) | $dH_{c2}/dT$ (T=$T_c$)(T/K) | Maki parameter | $H_{c2}$ (T=0)(T) | $\xi$ (nm) | $v_F$ ($10^5$ m/s) |
|---|---|---|---|---|---|---|---|
| 3 | 225 | 105 | -0.69 | 0.364 | 46.9 | 2.6 | 2 |
| 5 | 310 | 167.7 | -0.67 | 0.352 | 73.0 | 2.1 | 2.6 |

**Summary**

In conclusion, we have successfully synthesized the new $CaMg_2$-based ternary superhydride with *fcc* symmetry. It exhibits high superconducting transition temperature $T_c$ of ~168 K at 310 GPa, which is lower than the 1:1 CaMg-based superhydride at comparable pressures. Our results provide the first example of the synthesis of the novel ternary superhydrides using the existing binary alloys as starting reactants.

**Methods**

The Ca-Mg based superhydrides were synthesized from $CaMg_2$ intermetallic (99%, purchased from ACI Alloys Inc), 1:1 Ca-Mg alloy (synthesized using arc melting technique) and ammonia borane (AB, 97%, Sigma-Aldrich) at megabar pressures. The elemental composition of the purchased alloys was checked by energy dispersive X-ray analysis (EDS) method to ensure its homogeneity. Pressure was generated by custom diamond anvil cells (DAC). The anvils have typical culet sizes ~18-25 μm and were bevelled at 8° to the diameters of ~250-300 μm. Pressure was calibrated by the edge of diamond anvils' Raman spectra.[48, 49] One side of the anvils was sputtered with tantalum leads which were then covered with gold in the four-probe van der Pauw geometry. The mixtures of epoxy and *c*-BN, MgO or CaO were used to insulate the T301 stainless steel gasket and electrodes. The insulating gasket was pre-indented to ~8.0 μm thickness where upon a hole was laser-drilled with the diameter comparable to the culet size.

Sample loading were conducted in the argon-filled glovebox with the $O_2$ and $H_2O$ level lower than 0.1 ppm. Initially, transparent AB polycrystalline crystals were loaded in the ~8.0 μm thickness gasket chamber and pre-compressed to ~3.5 GPa. Then a small ~2.0 μm thick piece of CaMg$_2$ or 1:1 CaMg alloy was then placed on top of the four electrodes of the upper diamond. The sample size is generally half of the culet diameter. The loading was initially compressed to ~5.0 GPa in the glovebox and the pressure further increased to desirable value after the DACs were transferred outside. For the high temperature and high-pressure synthesis, the cylinder side of anvils was heated by a pulsed YAG laser with the spot size of ~10 μm. To avoid diamond failure, samples were heated gradually till obvious glowing was observed (~1500 K). During heating, the laser spot was carefully moved horizontally and vertically to ensure the complete chemical reaction of the CaMg$_2$ (CaMg) and AB.

Electrical transport measurements were conducted by the van der Pauw technique using a Keithley 6220 current source and a 2000 multimeter. The DAC was inserted in the liquid nitrogen cryostat and cooled from 296 to 80 K. The temperature was recorded with an accuracy of ~1.0 K with the aid of a silicon diode thermometer which was attached close to the sample. The electrical current ranged between 100 μA and 0.5 mA. After cooling to the base temperature of ~80 K, resistance data were recorded during the slow warm up at a rate of ~0.2 K/min. Magnetotransport measurements of Sample 3 and Sample 5 were performed in nonmagnetic DACs under external fields of 0-14 T in a Quantum Design's Physical Phenomena Measurement System (PPMS). The magnetoresistance was recorded by the PPMS AC resistance bridge and by the LakeShore Cryotronics 370 resistance bridge.

Most of high pressure X-ray diffraction measurements were performed at beamline 13-ID-D, GSRCARS of the Advanced Photon Source (APS), Argonne National Laboratory (ANL). The X-ray wavelength λ is 0.2952 and 0.3344 Å. Part of synchrotron X-ray data were collected at Beamline P02.2 at PETRA III, Deutsches Elektronen-Synchrotron (DESY) using beam spot size of ~2×2 μm$^2$ and a 2D area Perkin Elmer XRD 1621 detector ($\lambda$ = 0.2902 Å). Clean XRD patterns were obtained by subtracting the background as described elsewhere.[44] All diffraction images were integrated using the *Dioptas* software.[50]

**Acknowledgements**

M. I. E. is thankful to the Max Planck community for the support, and Prof. Dr. U. Pöschl for the constant encouragement. W. Cai would like to acknowledge financial support from National Natural Science Foundation of China (No. 12274062) and the Natural Science Foundation of Sichuan Province (No. 2022NSFSC0297). The authors would like to acknowledge Dr. A. P. Drozdov for his experimental assistance. The X-ray experiments of this research were carried out at 13 ID-D beamline, GeoSoilEnviro CARS (The University of Chicago, Sector 13), Advanced Photon Source (APS), Argonne National Laboratory. GeoSoilEnviro CARS is supported by the National Science Foundation-Earth Sciences (EAR-1634415) and Department of Energy-GeoSciences (DE-FG02-94ER14466). This research used resources of the Advanced Photon Source, a U.S. Department of Energy (DOE) Office of Science User Facility operated for the DOE Office of Science by Argonne National Laboratory under Contract No. DE-AC02-06CH11357. The National High Magnetic Field Laboratory is supported by the National Science Foundation through NSF/DMR-1644779 and the State of Florida and the U.S. Department of Energy.

# Supplementary Information

# Superconductivity above 180 K in Ca-Mg Ternary Superhydrides at Megabar Pressures

Weizhao Cai[1,2], Vasily S. Minkov[2], Ying Sun[3], Panpan Kong[2], Krista Sawchuk[4], Boris Maiorov[4], Fedor F. Balakirev[4], Stella Chariton[5], Vitali B. Prakapenka[5], Yanming Ma[3], Mikhail I. Eremets[2,*]

[1]*School of Materials and Energy, University of Electronic Science and Technology of China, Chengdu 611731, Sichuan, China*

[2]*Max Planck Institute for Chemistry, Hahn Meitner Weg 1, Mainz 55128, Germany*

[3]*State Key Laboratory of Superhard Materials and International Center for Computational Method & Software, College of Physics, Jilin University, Changchun 130012, China*

[4]*MagLab, Pulsed Field Facility, Los Alamos National Laboratory, Los Alamos, NM 87545, USA*

[5]*Center for Advanced Radiation Sources, University of Chicago, Chicago, IL 60637, USA*

*E-mail: m.eremets@mpic.de

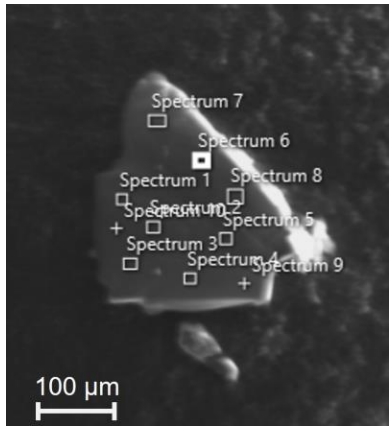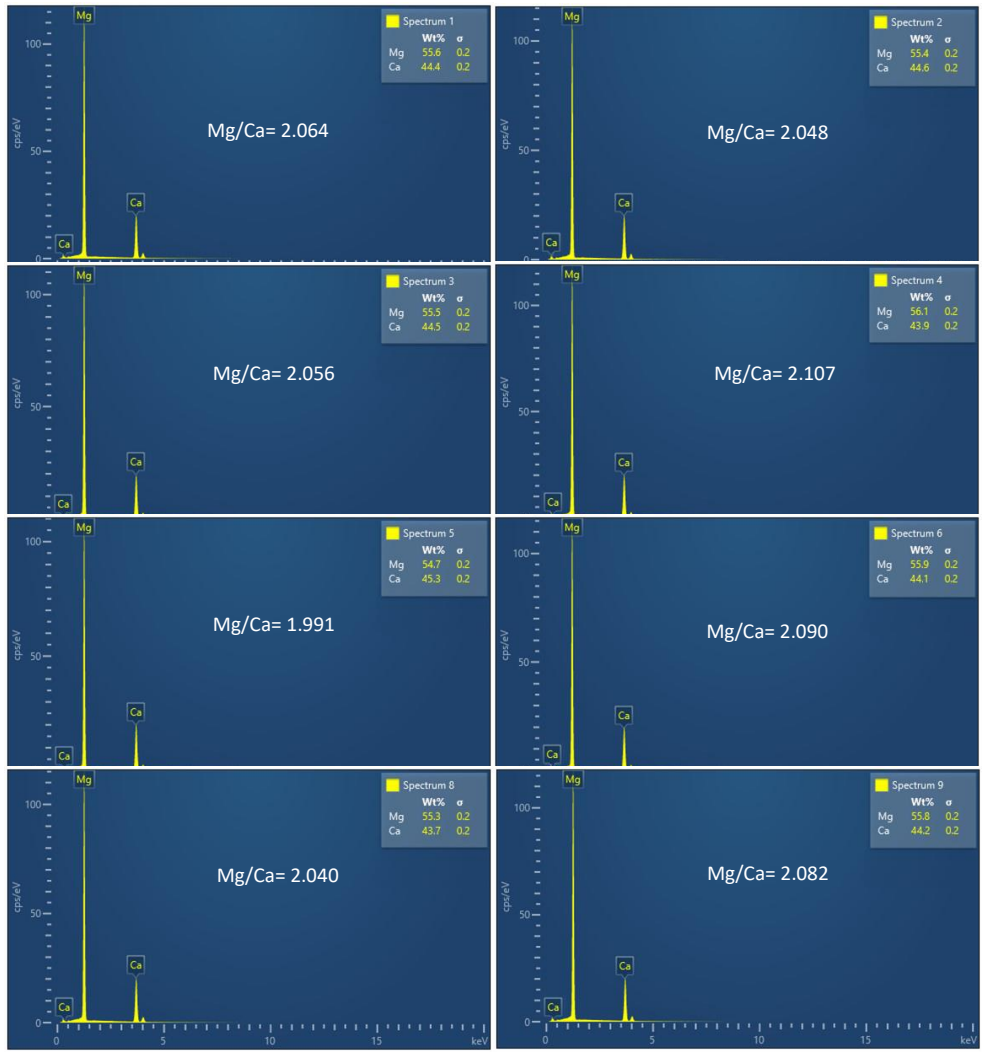

**Figure S1**. SEM image and selective EDS spectra of the CaMg$_2$ alloy taken from different positions.

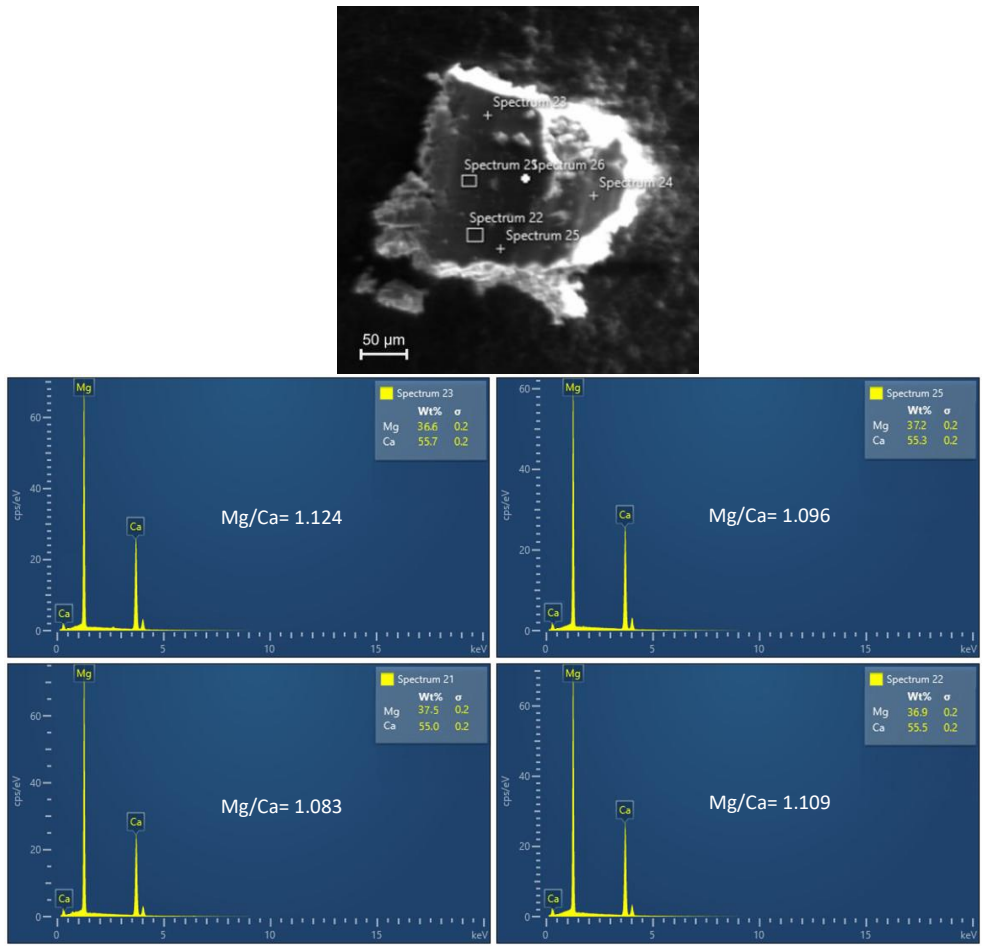

**Figure S2**. SEM image and representative EDS spectra of the 1:1 Ca-Mg alloy taken from different positions.

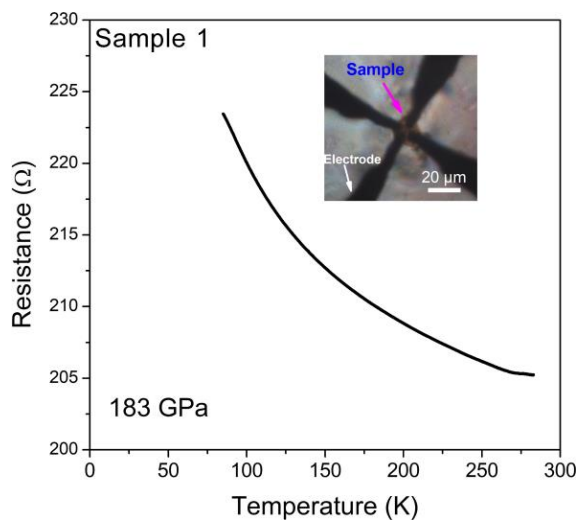

**Figure S3**. The electrical resistance of the synthesized Sample 1 as a function of the temperature. The data are collected from three-probe electrical resistance measurement. The semitransparent sample is shown in the inset with the image taken in the transmission light.

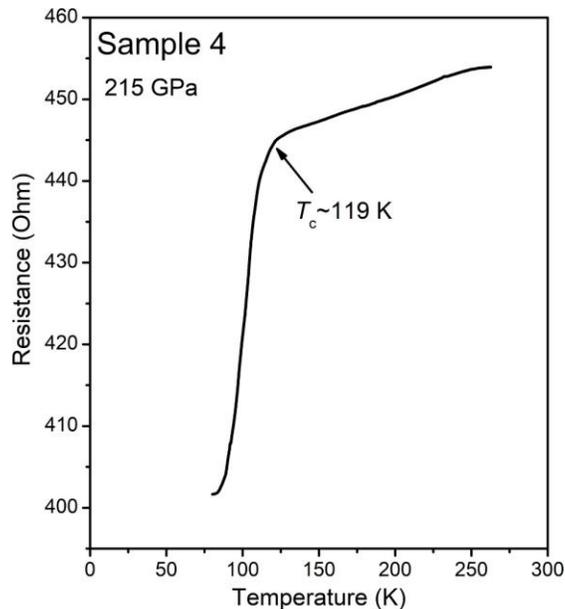

**Figure S4**. The observation of superconductivity in Sample 4. Both samples were synthesized from the mixture of $CaMg_2$ alloy and ammonia borane. (a) Three-probe electrical resistance measurement of Sample 4 at 215 GPa.

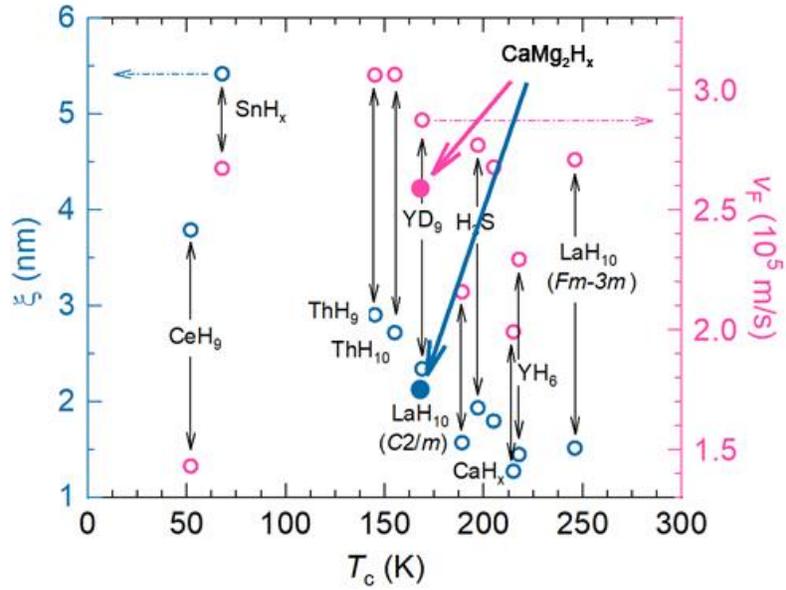

**Figure S5. Distribution of coherence lengths and Fermi velocities in relation to superconducting transition temperature in high-$T_c$ hydrides.** The plot is reproduced form Reference (Sun D, *et al.* High-temperature superconductivity on the verge of a structural instability in lanthanum superhydride. *Nat Commun* **12**, 6863 (2021).) with the addition of the values for $CaMg_2H_x$ reported in this work.

**Table S1.** List of samples synthesized at different pressures

| DAC | Images | Synthesis details | Electrical resistance measurements | X-ray diffraction data |
|---|---|---|---|---|
| Sample 1 | 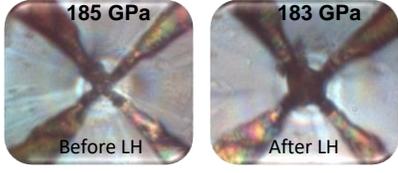 185 GPa Before LH / 183 GPa After LH. 25 μm diameter culet | CaMg$_2$ and ammonia borane alloy was compressed to 185 GPa at room temperature and laser-heated at 1200-1600 K, pressure was dropped to 183 GPa. | Three probe electrical resistance data. The synthesized sample shows semiconductor character down to 80 K. | Almost pure *fcc* phase: Lattice parameter: $a = 4.6995(3)$ Å, $V = 103.79(2)$ Å$^3$ |
| Sample 2 | 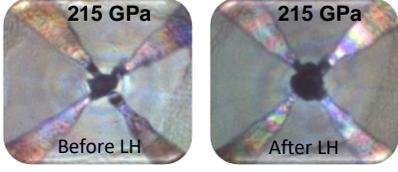 215 GPa Before LH / 215 GPa After LH. 25 μm diameter culet | CaMg$_2$ and ammonia borane mixture was compressed to 215 GPa at room temperature and laser-heated at ~1600 K. | Four probe electrical resistance data. $T_c = 121(1)$ K (215 GPa); $T_c = 132(2)$ K (230 GPa); $T_c = 139(2)$ K (246 GPa). Zero resistance was observed. Purple squares in Figure 3c. | At 230 GPa, the main phase was identified as the *fcc* phase with lattice parameter: $a = 4.5173(1)$ Å, $V = 92.18(1)$ Å$^3$. |
| Sample 3 | 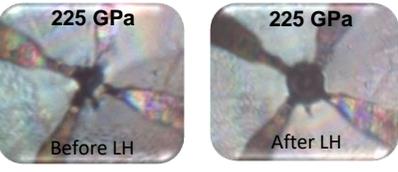 225 GPa Before LH / 225 GPa After LH. 23 μm diameter culet | CaMg$_2$ and ammonia borane was compressed to 225 GPa at room temperature and laser-heated at ~1700 K. | Four probe electrical resistance data. $T_c = 105(2)$ K (225 GPa). Yellow diamonds in Figure 3c. | Almost pure *fcc* phase: Lattice parameters 218 GPa: $a = 4.5893(2)$ Å, $V = 96.65(1)$ Å$^3$. 242 GPa: $a = 4.4969(2)$ Å, $V = 90.94(1)$ Å$^3$. |
| Sample 4 | 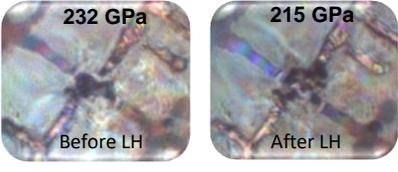 232 GPa Before LH / 215 GPa After LH. 20 μm diameter culet | CaMg$_2$ and ammonia borane mixture was compressed to 232 GPa at room temperature and laser-heated at 1200 K. Pressure was dropped to 215 GPa. Further compression led to the diamond failure. | Since one electrode broke during laser heating (LH), three probe electrical resistance data were collected. $T_c = 119(2)$ K (at 215 GPa). Blue circles in Figure 3c. | n/a |
| Sample 5 | 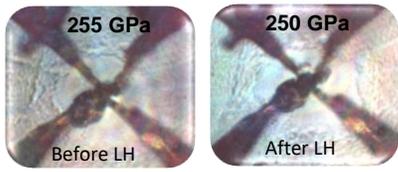 255 GPa Before LH / 250 GPa After LH. 18 μm diameter culet | CaMg$_2$ and ammonia borane was compressed to 255 GPa at room temperature and laser-heated at 1500 K, pressure was dropped to 250 GPa. And the pressure was increased to 285 and 310 GPa at room temperature. | Four probe electrical resistance data. $T_c = 143(2)$ K (250 GPa); $T_c = 153(2)$ K (285 GPa); $T_c = 168(2)$ K (310 GPa). Black squares in Figure 3c. | n/a |
| Sample 6 | 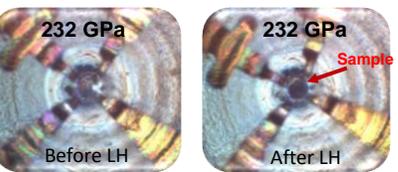 232 GPa Before LH / 232 GPa After LH (Sample). 20 μm diameter culet | 1:1 Ca-Mg alloy and ammonia borane were compressed to 232 GPa at room temperature and laser-heated at ~1600 K, and pressure was increased to 248, 280, and 321 GPa, respectively. Further compression led to the diamond failure. | Four-probe electrical resistance data $T_c = 152(2)$ K (232 GPa); $T_c = 165(2)$ K (248 GPa); $T_c = 174(2)$ K (280 GPa); $T_c = 182(2)$ K (321 GPa). Green diamonds in Figure 3c. | n/a |

Marked red: DAC used for magnetic measurements.